# Opinion Leader Detection in Online Social Networks Based on Output and Input Links

Zahra Ghorbani[1]. Seyed Hossein Khasteh[2]. Saeid Ghafouri[3]

**Abstract** The understanding of how users in a network update their opinions based on their neighbours' opinions has attracted a great deal of interest in the field of network science, and a growing body of literature recognises the significance of this issue. In this research paper, we propose a new dynamic model of opinion formation in directed networks. In this model, the opinion of each node is updated as the weighted average of its neighbours' opinions, where the weights represent social influence. We define a new centrality measure as a social influence metric based on both influence and conformity. We measure this new approach using two opinion formation models: (i) the Degroot model and (ii) our own proposed model. Previously published research studies have not considered conformity, and have only considered the influence of the nodes when computing the social influence. In our definition, nodes with low in-degree and high out-degree that were connected to nodes with high out-degree and low in-degree had higher centrality. As the main contribution of this research, we propose an algorithm for finding a small subset of nodes in a social network that can have a significant impact on the opinions of other nodes. Experiments on real-world data demonstrate that the proposed algorithm significantly outperforms previously published state-of-the-art methods.

## 1 Introduction

Recently, a large number of research studies have been undertaken to achieve a better understanding of individuals' everyday opinions on various subjects. This increase in attention is due to the fact that opinions crucially shape human behaviour in important areas such as economics, political science, cultural problems etc. Social networks play a fundamental role in the forming and reshaping of public opinions, and have therefore attracted a great amount of attention in the area of opinion formation studies (Duan, et al. 2014; Karlsen 2015; Ma and Liu 2014; Van den Brink, et al. 2013; Zhang, et al. 2014). Consider a prominent application in a viral marketing campaign that aims to convert a small number of initial users in order to eventually influence the majority of people in the market. This problem can be viewed as an example of finding "opinion leaders" in a social network, which is also the main challenge of this research paper. A further example might involve the forecasting of the final opinion of each node and the maximisation of the spread of influence via a network (Guo, et al. 2013; Hao, et al. 2012; Jiang, et al. 2011). Society can also be seen in terms of a network, and we can therefore apply

✉ Zahra Ghorbani
z.ghorbani@mail.kntu.ac.ir

Seyed Hossein Khasteh
khasteh@kntu.ac.ir

Saeid Ghafouri
Saeidghafouri@email.kntu.ac.ir

[1] Department of Computer Engineering
K. N. Toosi University of Technology
Tehran, Iran

[2] Department of Computer Engineering
K. N. Toosi University of Technology
Tehran, Iran

[3] Department of Computer Engineering
K. N. Toosi University of Technology
Tehran, Iran

our approach for finding opinion leaders to a large number of social problems such as control of the spread of disease (Miller and Kiss 2014), game theory based on financial problems (Simko and Csermely 2013) and any other problems related to finding the influential nodes in an arbitrary network (Bao, et al. 2017; Fei, et al. 2017; Liu, et al. 2016; Loeper, et al. 2014; Malliaros, et al. 2016; Mo, et al. 2015; Srinivas and Velusamy 2015; Wang, et al. 2017).

We should emphasise that our method should not be confused with influence maximisation research studies that operate under certain propagation models (Kempe, et al. 2003). Our research paper finds opinion leaders using opinion formation models, which despite certain similarities has critical differences from influence maximisation studies. Firstly, influence maximisation studies are based on the solution of an optimisation problem through leveraging some kind of greedy algorithm, and although the details of these vary from one research study to another, they all use the same technique, as mentioned above. However, our research paper takes a different approach involving the use of an iterative algorithm. This paper aims to find the values of a concept called "centrality" for each node in order to use these further in finding opinion leaders. In addition, we test opinion formation models. In most influence maximisation studies, each node has two Boolean values. A node is either active or inactive, and unlike opinion formation models, does not have continuous values. We explain our opinion formation model in more detail in Section 2.3.

The main approaches to opinion formation theory can be divided into methods based on (i) *Bayesian learning* and (ii) *non-Bayesian learning*. Opinion formation based on Bayesian updates, such as the work presented in (Buechel, et al. 2015b; Hudli, et al. 2012; Jia, et al. 2015; Kaiser, et al. 2013; Loeper, et al. 2014; Van den Brink, et al. 2013) involves updating the opinion of each user based on the Bayes theorem. The difficulty with Bayesian methods is that they generally form and update opinions based on the underlying state of and signals from the world, and consequently the results obtained by such methods cannot be easily generalised to other states of the network. This does not mean that these methods are not at all generalizable; however, compared to non-Bayesian methods, they are much less generalizable. Opinion formation methods based on non-Bayesian approaches employ rule-of-thumb methods to form the opinion of agents, giving good approximations of each agent's behaviour and avoiding the difficulties of Bayesian approaches. These approaches are widely applied in the literature, such as the research studies presented in (Cho, et al. 2012; Duan, et al. 2014; Dubois and Gaffney 2014; Eom and Shepelyansky 2015; Kandiah and Shepelyansky 2012; Karlsen 2015; Loeper, et al. 2014; Ma and Liu 2014; Mor and Girard 2011; Xiao and Xia 2010; Zhang, et al. 2014); these are closely related to the Degroot model (DeGroot 1974), and form the foundations of our proposed opinion formation model. Previous researchers in the area of opinion formation studies have typically focused on undirected networks, and have assumed no significant distinction between ingoing and outgoing links in the network. In this paper, we introduce a new model for directed networks in which are two type of links for each node: (i) ingoing; and (ii) outgoing. Outgoing links represent connections between that node and its followers, while ingoing links are the connections to its followees[4]. In our model, the opinion of a node is updated based on the weighted sum of the opinions of its neighbour nodes; the weights of these neighbour nodes are given by their centrality, which we will define in Section 2. The main component of this research is based on the concept of centrality, and we design an efficient algorithm for finding the most influential nodes, that is, nodes that impose their own opinion on a significant proportion of the other nodes. Previous centrality models, apart from traditional models such as betweenness or degree centrality, do not consider both ingoing and outgoing links, and thus are not optimally suitable for finding opinion leaders. For example, game centrality aims to find the most suitable nodes to defect in a network modelled with game theory, and can be modelled by a game (Simko and Csermely 2013). Further examples of game theory-based approaches are Shapley centrality (Chen and Teng 2017), leverage centrality, the main focus of which is on the specific domain of brain networks (Joyce, et al. 2010), and perturbation centrality, which is a model inspired by vessel communications (Szalay and Csermely 2013); (Candeloro, et al. 2016). The main application of this centrality is in animal behaviour prediction. Another example of these domain-specific centralities is functional centrality, which is related to protein interaction networks (Tew, et al. 2007). Some works have even expanded their criteria to multilayer networks (Wang and Zou 2018) or applications of centrality in aspects of daily life such as urban transport (Wang and Fu 2017). There are also comprehensive surveys (Das, et al. 2018) dedicated to introducing the most recent works in the area of centrality. All of these methods perform well within their specific domains or other networks. However, they are not conceptually suitable for finding opinion leaders; this is true even for novel models like k-shell centrality, which shows very good great performance in various domains (Kitsak, et al. 2010). In addition, the centrality measures used in other papers to find opinion leaders are specifically designed for the opinion formation networks (e.g. (Buechel, et al. 2015a; Dubois and Gaffney 2014; Eom and Shepelyansky 2015; Jia, et al. 2015). We are not the first to consider opinion networks in our centrality model, since previous works such as (Cho, et al. 2012) consider opinion networks in their centrality models, but we choose a different approach to do so. In all the networks considered here, the experimental results indicate that our proposed method significantly outperforms the other algorithms, such as those based on eigenvector, betweenness and closeness centralities. Section 2 of this paper will discuss methods and materials, while Sections 3 and 4 present the experimental results and the conclusion and future research respectively.

---

[4] Other nodes which follow that particular node.

## 2  Why another model?

As mentioned in the introduction section, there are already a number of state-of-the-art methods dedicated to finding influential nodes and opinion leaders, in which more central nodes are referred to under the name of "centrality". Despite the strong performance achieved by most of these methods, we believe that there are a number of deficiencies that can provide good motivation for seeking a new novel method. Firstly, although there are plenty of research studies dedicated to the problem of finding opinion leaders, their numbers are still lower than the number of articles about influence maximisation. Another reason is that some of the existing research studies are domain-specific, and although their methods show good performance within these specific domains, their performance in other areas such as opinion formation is not guaranteed.

From a technical point of view, we believe that our method has the following advantages: (i) conformity and input and output links are considered in our computations; (ii) a heuristic is proposed to reduce the number of computations in each step; and (iii) a novel opinion formation model is proposed that takes the conformity of nodes into account.

Table 1 shows a comparison of several state-of-the-art methods with our proposed method in terms of three technical criteria.

Table 1
*Comparison between different methods*

| Advantage / Method | (1) | (2) | (3) |
|---|---|---|---|
| Our proposed method | ✓ | ✓ | ✓ |
| (Wang, et al. 2017) | ✗ | ✗ | ✗ |
| (Salehi-Abari and Boutilier 2014) | ✗ | ✓ | ✗ |
| (Bao, et al. 2017) | ✗ | ✓ | ✗ |
| (Buechel, et al. 2015b) | ✓ | ✗ | ✓ |

## 3  Methods and Materials

The main goal of this paper is to solve the problem of measuring the social influence of users in online social networks; this in turn is used as a metric to find the nodes. In real-world problems, the term "social influence" could refer to individuals' income, persuasiveness of speech, level of education or even their physical attributes, their power among other society ranking members, or the place of each individual in such rankings. Most of the existing research studies (Borgs, et al. 2014; Eom and Shepelyansky 2015; Jia, et al. 2017; Jia, et al. 2015; Loeper, et al. 2014; Van den Brink, et al. 2013), measure social influence by raising the question of the extent to which a node can affect other nodes of the network. We extend these methods by taking into account another attribute of nodes called conformity. We consider conformity to be a metric of the bias of a node's opinion based on the opinions of the other nodes in the network. As a result, our approach improves upon previous methods by not only considering a node's influence but also using its conformity to evaluate the extent to which a node adheres to its current opinions. For example, consider the Instagram network, in which ingoing links represent the connections to the users that are followed, and outgoing links represent the connections from the user's followers. In citation networks of scientific papers, the ingoing links represent connections among papers that are cited by a given paper, while the outgoing links represent connections to papers that have cited that paper. The conformity of each node is based on a number of parameters, such as the number of its input links, the stability of each node's opinion, the nature of the specific relations in the social network and so forth. All of these parameters should be considered in calculating the conformity of each node; however, to simplify our formulation, we simply omit other factors and consider only the number of input links as a measure of conformity.

### 3.1 Existing metrics

The discovery of influential nodes is a challenging issue in the field of complex network analysis. A number of large cross-sectional studies suggest that heuristic methods can efficiently recognise the most influential nodes of networks. The influence of a node is primarily associated with the mathematical definition of a quantity called "centrality", which refers to several measures that define each node's importance in a graph. As shown in the next section, we compare our proposed method for identifying influential nodes using several other measures of centrality such as degree, betweenness, closeness and eigenvector centralities. The simplest measure of centrality is degree centrality, which denotes the number of immediate neighbours of a node. Degree centrality can be interpreted as the number of walks of length one, starting from that particular node, and this measures the local influence of a node. The degree centrality for a directed graph has one of two forms: either the in-degree, which counts the number of direct ties to the node, or the out-degree, which counts the number of ties that a node directs to others. Each of these forms has a different meaning in networks. For comparison purposes, we use out-degree centrality here, in which a node with high out-degree indicates an influential node.

The betweenness centrality (Freeman 1977) is defined as a fraction of the shortest paths between each pair of nodes passing through the considered node. Equation 1 shows the betweenness centrality of node $i$:

$$g_i = \sum_{j \neq k \neq i} \frac{\sigma_{jk}(i)}{\sigma_{jk}} \tag{1}$$

where $\sigma_{jk}$ refers to the total number of shortest paths from node $j$ to node $k$, and $\sigma_{jk}(i)$ denotes the number of these paths passing through $i$. Another measure of centrality is closeness centrality, defined as the inverse of the average of the shortest-path distance to other nodes. In this measure, nodes with high closeness reach others quickly and play an important role in spreading information. Equation 2 reveals the closeness centrality for a given node $i$ in a directed network as:

$$C_i = \frac{N-1}{\sum_{j=1}^{N} d(i,j)} \tag{2}$$

Eigenvector centrality, as introduced by Bonacich (Bonacich 1987), links nodes to other important nodes, known as central nodes. It defines the centrality of a node based on the centrality of the other nodes connected to it. For a node $i$ with neighbouring nodes $N(i)$, the centrality score can be calculated as follows:

$$c_i = \frac{1}{\lambda} \sum_{j \in N(i)} c_j = \sum_j A_{ij} c_j \tag{3}$$

or in matrix format,
$$AC = \lambda C \tag{4}$$

In general, there are many different eigenvalues where $\lambda$ refers to a non-zero eigenvector solution. Each row in $C$ represents a node's eigenvector centrality.

Another centrality that is often used for comparison is the PageRank centrality (Brin and Page 2012; Perra and Fortunato 2008), which is a variant of eigenvector centrality. It was introduced for use by the Google web search engine to rank web pages in the World Wide Web (Brin and Page 2012). It ranks web pages based on the sum of the ranks, and can be computed through the following equation:

$$PR(i) = (1-d) + d \sum_{j \in A_{i,in}} \frac{PR(j)}{d_j^{out}} \qquad (5)$$

where $N$ is the total number of pages on the Web, $d$ is a damping factor and $d_j^{out}$ is the out-degree of $j$. PageRank centrality is currently used to rank nodes in various types of directed networks (e.g. its usage in designing an opinion formation model) (Eom and Shepelyansky 2015). Frahm et al. (2014) have also tried to model the Google citation network by adopting the PageRank algorithm.

### 3.2 A new centrality measure: Global centrality

Centrality is a well-known measure in network theory and has several variations in literature reviews (Fei, et al. 2017; Liu, et al. 2016; Srinivas and Velusamy 2015; Wang, et al. 2017). Centrality is defined as the main indicator of the extent to which a node is influential.

In this section, we propose a new centrality measure for finding the top $k$ influential nodes of the network. We first define a new parameter called the "effective degree" for each node.

**Definition 1** *The effective degree of a node $i$ is calculated as follows:*

$$EDeg(i) = outdegree(i) - indegree(i) \qquad (6)$$

We also define another parameter, which we call the degree centrality. This parameter measures the local social influence, and its value for each node is calculated based on the sum of that node's effective degree, plus the effective degree of the nodes.

**Definition 2** *The degree centrality of node $i$ is calculated as follows:*

$$C_i = EDeg_i + \sum_{j \; out-neighbours \; of \; i} EDeg_j \qquad (7)$$

We prefer to work with smaller values and for all the calculated degree centralities to fall within a positive interval; we therefore subtract the minimum degree centrality that exists in the network from the degree centrality value of each node:

$$C_i = C_i - min(C_j) \quad (j \in V) \qquad (8)$$

where $V$ refers to the set of all nodes in the network.

So far, all we have done has been the calculation of the centrality of nodes based on their immediate neighbours' degrees. However, we now propose a more generic metric that is able to find the centrality based not only on each node's immediate neighbours, but also the structure of the entire graph. In doing so, we define a recursive equation. We refer to the centrality obtained by this recursive equation as the global centrality. This is due to the fact that each node that has neighbours with higher centrality has higher centrality itself.

To calculate this global centrality measure, we first consider the graph $G = (V, E)$, in which $N = |V|$ and $L = |E|$ indicate the number of nodes and edges, respectively. We represent this graph with a $N \times N$ adjacency matrix $A$.

Each element in the matrix $A$ can either be one or zero; if $a_{ij}$ equals one, this indicates a direct connection between node $i$ and $j$, whereas a value of zero implies that no connection exists between these two nodes. We normalise the elements of each row by dividing them by the sum of the elements of their corresponding row, in order to transform the adjacency matrix into the form of a stochastic matrix[5]. We use the stochastic matrix property to prove that our model in Equation 17 converges to the true value.

$$W(i,j) = \frac{A(i,j)}{\sum_k A(i,k)} \qquad (9)$$

Obviously, the sum of each row of matrix $W$ is equal to one, and thus the matrix is in the form of a stochastic matrix.

Using this stochastic matrix, we propose a measure for computing the global centrality. Unlike the degree

---

[5] A matrix in which the values of each of the rows sum to one.

centrality in Equation 7, the stochastic matrix considers the structure of the entire graph.

$$C^g = \alpha C + (1 - \alpha)WC^g \qquad (10)$$

Or in another form

$$C^g = (I - (1 - \alpha)W)^{-1}\alpha C \qquad (11)$$

In these equations, $C$ represents the local centrality, or what we have previously defined as the degree centrality in Equation 7. The parameter $\alpha$ refers to a localisation parameter that optimises the extent to which $C$ and $WC^g$ influence each node's global centrality. $\alpha$ is the value that controls the tradeoff between the global and the local views of the graph. If the value of $\alpha$ is close to one, our computation will be closer to the local centrality, and if it is close to zero, the global centrality will be mainly based on the global view of the whole graph.

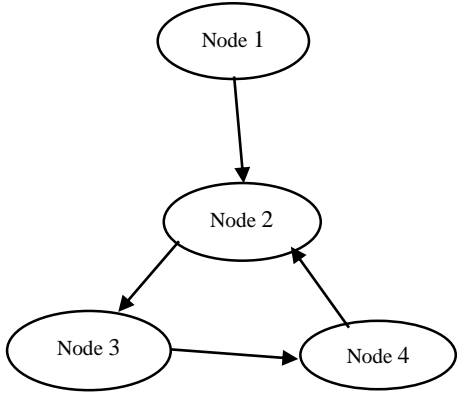

*Figure 1. A sample graph for comparison of centrality measures*

We compare our proposed method with existing metrics in order to find the influential nodes. We also introduce an opinion formation model and show that our method can find opinion leaders in the network more effectively and can speed up the process of spreading ideas. We illustrate the performance of our method compared to other methods using a simple example. Consider the graph shown in Figure 1: node 1 influences the other nodes, but does not follow any other node. For this reason, it is considered to be the best candidate for spreading an idea across the network. We implemented and discussed the centrality measure algorithms for this graph, and the results are presented in Table 1. As can be observed from Table 2, our method idenitifies node 1 as a leader, while other methods identify node 2 as a leader. The Cg column denotes our proposed algorithm.

Table 2
*Comparison between centrality measures in a sample graph[6]*

| NodeID | Closeness | Betweenness | Eigenvector | PageRank | Cg |
|---|---|---|---|---|---|
| Node 1 | 0.6 | 0 | 0.32 | 0.11 | 1.0 |
| Node 2 | 1.0 | 1.0 | 1.0 | 1.0 | 0.82 |
| Node 3 | 0.75 | 0 | 0.67 | 0.96 | 0.89 |
| Node 4 | 0.75 | 0 | 0.67 | 0.9 | 0.7 |

### 3.3 The Degroot opinion formation model

In this section, we present a brief review of the Degroot model (DeGroot 1974). We assume that time is discrete $t = 0, 1, 2, ...$ and that each agent initially has a predefined opinion $x_i$ concerning some topic. The opinions are values between zero and one, i.e. $x_i \in [0,1]$. The opinions of all agents at time $t$ are collected in $x(t) \in R$ (we assume $n$ agents). In each period, agents talk to each other, and finally each of the agents updates its own opinion by obtaining a weighted average of the opinions of its neighbours. We then have:

$$x_i(t + 1) = \sum_j T_{ij} x_j(t) \qquad (12)$$

where the weight matrix $T = [T_{ij}]_{i,j \in N}$ represents the level of truth that any agent places on the opinion of any other agent. All the elements of $T$ satisfy $t_{ij} \in [0,1]$ for $0 \leq i, j < n$. The elements in each row of the matrix $T$ sum to one, which implies that it is row stochastic. Another variation of this formula can be expressed as:

$$x(t + 1) = Tx(t) = T^{t+1}x(0) \qquad (13)$$

An interesting variation of this model was developed by Friedkin and Johnsen (Friedkin and Johnsen 1990; Jia, et al. 2015). Their model addresses opinion formation under social influence and assumes that an agent $i$ adheres to its initial opinion with a certain probability $g_i$, and that the agent is socially influenced by the other agents with probability $1 - g_i$. According to the classical model, we have:

$$x_i(t + 1) = g_i x_i + (1 - g_i)(a_{i1}(t) + \cdots a_{in} x_n(t))$$

$$(14)$$

---
[6] Values are normalised.

or in matrix notation,
$$x(t+1) = Gx(0) + (I-G)Ax(t) \quad (15)$$

Here, $G$ is the diagonal matrix where $0 < g_i < 1$ on the diagonal and $I$ is the identity matrix.

## 3.4 Proposed opinion formation model

In this section, we present a brief overview of the opinion formation model that we intend to use in our benchmarking alongside the Degroot opinion formation model. This model has a different structure from the opinion-leader-finding algorithms, with a different definition of global centrality. In other words, this opinion formation model is one of the novel aspects of this research. Consider a network consisting of $n$ agents (nodes) in which the agents interact with each other. Each agent holds certain opinions on specific topics, and their opinions are updated on a regular basis based on their neighbours' opinions. The issues of how to update opinions and in which situations each agent's opinion is updated are referred to as the opinion formation model in the literature.

Opinion formation models that have been presented so far are either applicable only to undirected graphs (Jalili 2013; Mor and Girard 2011) or do not take into account the differences between ingoing and outgoing links (Buechel, et al. 2015b; Chen, et al. 2016; Eom and Shepelyansky 2015; Jia, et al. 2015; Kandiah and Shepelyansky 2012). In this section, we present a new opinion formation model that takes into account the differences between ingoing and outgoing links.

Consider a graph with $n$ nodes in which each node corresponds to an agent and each neighbourhood relation between two nodes is represented by an edge. For example, $i$ is $j$'s neighbour if and only if a directed edge $(i,j)$ exists. We define a weight matrix $W$, and use this matrix to calculate the new opinion of each node. Its elements indicate how much influence each node experiences from its neighbours. The opinion of agent $i$ is represented as $x_i$ and is updated by a linear combination of its neighbours' opinions. It can be calculated as follows:
$$x_i = d_{ii}x_i(0) + \sum_{j=1, j\neq i}^{n} w_{ij}x_j \quad (16)$$

In matrix form, we have:

$$X = WX + DX \circ \quad (17)$$

or in another form:
$$X = (I-W)^{-1}DX \circ \quad (18)$$

As explained earlier, $W$ refers to the influence matrix, in which the elements on the main diagonal are all zero and the other elements represent each neighbour's influence on a node. Matrix $D$ is a diagonal matrix in which the elements of the main diagonal represent the level of coherence of each agent regarding its own opinion.

The main problem in this model is how to set the values of the influence matrix $W$. In this research study, we assume that each node has an influence proportional to its degree centrality, which is calculated using the following formula:
$$d_{ii} = \frac{C_i}{\sum_{j \in in-neighbours\ of\ i} C_j + C_i} \quad (19)$$

$$w_{ik} = \frac{C_k}{\sum_{j \in in-neighbours\ of\ i} C_j + C_i} \quad (20)$$

In cases where we have a Eulerian graph,[7] which has equal in-degree and out-degree for each node, we set the corresponding $C_i$ zero values to a value of $1/100$ to ensure that the matrix $(I-W)^{-1}$ is invertible.

Our opinion formation model is a recursive model, and we build the opinion of each node based on the opinion of other nodes. In Section 2.3.1, we prove that this equation converges to a unique answer. In Section 2.3.2, we presented a method of finding this unique answer.

### 3.4.1 Solving the opinion formation model

Equation 17 has the form of a system of linear equations, and we can solve this by various direct and iterative methods. Cramer, Gaussian elimination, Gaussian Jordan elimination, and inverse matrix computation are examples of direct methods that are not suitable for problems with large numbers of variables due to their high computational complexities. For example, the inverse matrix method has a time complexity of $O(n^3)$, and the Cramer method has a complexity of $O(n!)$, making these methods impractical for problems with more than

---
[7] Directed graphs in which every node has the same in-degree and out-degree.

three variables. We need to solve the following equation:

$$X = WX + DX° \qquad (21)$$

In addition, in matrix form, we have

$$\begin{bmatrix} x_1 \\ x_2 \\ \vdots \\ x_n \end{bmatrix} = \begin{bmatrix} 0 & \frac{w_{12}}{\Sigma_k w_{1k}} & \frac{w_{13}}{\Sigma_k w_{1k}} & \cdots & \frac{w_{1n}}{\Sigma_k w_{1k}} \\ \frac{w_{21}}{\Sigma_k w_{2k}} & 0 & \frac{w_{23}}{\Sigma_k w_{2k}} & \cdots & \frac{w_{2n}}{\Sigma_k w_{2k}} \\ \vdots & \vdots & \vdots & \ddots & \vdots \\ \frac{w_{n1}}{\Sigma_k w_{nk}} & \frac{w_{n2}}{\Sigma_k w_{nk}} & \frac{w_{n3}}{\Sigma_k w_{nk}} & \cdots & 0 \end{bmatrix} \begin{bmatrix} x_1 \\ x_2 \\ \vdots \\ x_n \end{bmatrix} +$$

$$\begin{bmatrix} \frac{w_{11}}{\Sigma_k w_{1k}} & 0 & 0 & \cdots & 0 \\ 0 & \frac{w_{22}}{\Sigma_k w_{2k}} & 0 & \cdots & 0 \\ \vdots & \vdots & \vdots & \ddots & \vdots \\ 0 & 0 & 0 & \cdots & \frac{w_{nn}}{\Sigma_k w_{nk}} \end{bmatrix} \begin{bmatrix} x_1^{(°)} \\ x_2^{(°)} \\ \vdots \\ x_n^{(°)} \end{bmatrix} \qquad (22)$$

Our research study is mainly performed on large graphs such as the Facebook and Twitter networks with high numbers of variables; we therefore need methods with a linear time complexity. In view of this difficulty, we can use iterative methods such as Jacobi and Gauss-Siedel methods, since these methods converge with linear time complexities. Here, we use the Jacobi iterative method. As a result, the iterative solution for each step of the algorithm is formulated as follows:

$$x_i^{(t)} = d_{ii} x_i^{(°)} + \sum_{j=1, j \neq i}^{n} w_{ij} x_j^{(t-1)} \qquad (23)$$

### 3.5 Finding opinion leaders: An algorithmic approach

The global centrality defined in the previous section forms the foundation of our algorithm. We use the algorithm schema proposed by (Salehi-Abari and Boutilier 2014) and apply our centrality measure to find the opinion leaders. We need to find the nodes with most influence, which is similar to the global centrality. The influence rate for each node is computed using a recursive equation based on its neighbours' influence rates. In this definition, nodes that are connected to more influential nodes are more influential themselves.

Based on the definition of global centrality presented in Equation 10, we define the influence computation formula as follows:

**Definition 3** *The influence of node $i$ is calculated as follows:*

$$Influ_i = \alpha C_i + (1 - \alpha) \sum_{j \neq i} w_{ij} Influ_j \qquad (24)$$

or in matrix form:

$$Influ = \alpha C + (1 - \alpha) W Influ \qquad (25)$$

or in another form

$$Influ = (I - (1 - \alpha)W)^{-1} \alpha C \qquad (26)$$

where $Influ$ is the influence vector and $C$ is the degree centrality defined above in Equation 7. The iterative form of this equation in the $t + 1\,th$ step is

$$Influ_i^{(t+1)} = \alpha C_i + (1 - \alpha) \sum_{j \neq i} w_{ij} Influ_j^{(t)} \qquad (27)$$

Moreover, in matrix form, we have

$$Influ^{(t+1)} = \alpha C + (1 - \alpha) W Influ^{(t)} \qquad (28)$$

We set the initial influence value equal to the effective degree ($Influ^{(0)} = EDeg$), as introduced in Def. (6).

**Corollary 1** *Equation 27 has a unique solution and the iterative approach converges to a correct solution.*

*Proof:* In order to provide a proof for this, we need to prove that $1 - \alpha w$ is invertible. As explained above, $W$ is a row stochastic matrix, so the sum of the elements in each row sums to one and $\alpha < 1$. Therefore, $1 - \alpha w < 1$ and is invertible

In order to design an algorithm based on this approach, we define an error interval. We use this error interval to eliminate some of the nodes at each step of the algorithm that do not influence the final result. To proceed further in computing this interval, we need to use the definitions of norms, as follows.

**Definition 4** *A scalar norm is a type of norm that is defined for a vector or matrix and has the following properties:*

1. For each $x \in R^n \times n, \alpha \in R$ we have $\|\alpha x\| = |\alpha| \|x\|$
2. $\|XY\|_p \leq \|X\|_p \|Y\|_p$
3. $\|X^k\|_p \leq \|X\|_p^k$

**Definition 5** *An infinite norm is defined as follows:*

$$\|A\|_\infty = max_i(\sum_{j=1}^n |a_{ij}|) \qquad (29)$$

**Theorem 1** *The error value in the tth step of the proposed iterative method is bounded by the following interval:*

$$\|Influ - Influ^t\|_\infty \leq (1-\alpha)^t \|Influ - Influ^\circ\|_\infty \quad (37)$$

$Influ$ is the correct value of the influence, which is calculated by the proposed iterative method, and $Influ^\circ$ represents the initial value of the influence.

*Proof*: $Influ$ is the correct value of the influence after convergence of the proposed iterative approach. We therefore have:

$$Influ = (1-\alpha) W\, Influ \quad (30)$$

We also have

$$\|Influ - Influ^t\|_\infty = \|(1-\alpha)W\, Influ - (1-\alpha)W\, Influ^{(t-1)}\|_\infty \quad (31)$$

which is equal to:

$$= \|((1-\alpha)W)^t (Influ - Influ^\circ)\|_\infty \quad (32)$$

From Property 2 of Definition 4, we have:

$$\leq (1-\alpha)^t \|W^t\|_\infty \|Influ - Influ^\circ\|_\infty \quad (33)$$

In addition, from Property 3 of Definition 4, we can write:

$$\leq (1-\alpha)^t W_\infty^t \|Influ - Influ^\circ\|_\infty \quad (34)$$

By using the property of the infinite norm:

$$= (1-\alpha)^t (\max_j \sum_{j=1}^n |w_{ij}|)^t \|Influ - Influ^\circ\|_\infty \quad (35)$$

which yields the following result:

$$= (1-\alpha)^t \|Influ - Influ^\circ\|_\infty \quad (36)$$

Using this theorem, we can find the most influential nodes at each step.

**Theorem 2** *For each node in the tth iteration, the maximum value of the error is bounded by the following interval.*

$$|Influ_i - Influ_i^{(t)}| \leq (1-\alpha)^t (\max(Influ^\circ) - \min(Influ^\circ)) \quad (37)$$

*Proof*: From theorem 1, we have:

$$\|Influ - Influ^{(t)}\|_\infty \leq (1-\alpha)^t \|Influ - C\|_\infty \quad (38)$$

From the properties of the infinite norm:

$$\max_i |Influ_i - Influ_i(t)| \leq (1-\alpha)^t \max_i |Influ - C| \quad (39)$$

$$|Influ_i - Influ_i^{(t)}| \leq (1-\alpha)^t \max_i |Influ - Influ^{(\circ)}| \quad (40)$$

Matrix $W$ in Equation 27 is normalised. A node's influence in each iteration is therefore a linear combination of the normal coefficients of the influences of all nodes, and will never violate its maximum. Thus, we always have $\max(influ) \leq \max(influ^\circ)$. In addition, as a result of the equality of the initial influence and the degree centrality vector ($influ^\circ = C$), we have $\max(influ) \leq \max(C)$. We can therefore conclude:

$$\max_i |(Influ - Influ^\circ)| \leq \max(influ^\circ) - \min(influ^\circ) \quad (41)$$

$$|influ_i - influ_i^{(t)}| \leq (1-\alpha)^t (\max(C) - \min(C)) \quad (42)$$

Using the above theorem, we can conclude the following proposition:

**Proposition 1** *If $influ_i^{(t)} - influ_j^{(t)} > 2(1-\alpha)^t (\max(C) - \min(C))$ then $influ_i > influ_j$*

*Proof*:

$$influ_i^t - influ_j^t = influ_i^t + influ_i - influ_i - influ_j^t + influ_j - influ_j$$

$$= |influ_i - influ_i^t| + influ_i + |influ_j - influ_j^t| - influ_j \quad (43)$$

From Theorem 5, we have:

$$influ_i^t - influ_j^t \leq 2(1-\alpha)^t (\max(C) - \min(C)) + influ_i - influ_j \quad (44)$$

According to the proposition itself, we also know that:

$$influ_i^{(t)} - influ_j^{(t)} > 2(1-\alpha)^t(max(C) - min(C)) \quad (45)$$

With respect to this inequality, we know:

$$influ_i^t - influ_j^t \leq 2(1-\alpha)^t(max(C) - min(C)) + influ_i - influ_j \quad (46)$$

To make both of these inequalities true, we need:

$$influ_i - influ_j < 0 \quad (47)$$

We therefore have:

$$influ_i > influ_j. \quad (48)$$

Using this error interval, we can design a time-efficient algorithm for finding the top $k$ influential nodes in the network. In this algorithm, the initial influence of each node is set to the degree centrality, as discussed above. At each iteration, the value of the influence of each node is updated based on Equation 27, and the nodes that do not satisfy the conditions of Preposition (1) are eliminated.

The process of eliminating these nodes is as follows: in each iteration of the algorithm, we find the $k$th influential nodes which have a different in influence from the detected node that is greater than the threshold in Proposition (1); these are considered not to be influential nodes. The exclusion of less influential nodes in the current iteration is a heuristic used to reduce the computational complexity of the algorithm. We can be sure that these nodes are not in the set of the top $k$ most influential nodes in the current iteration, and there is therefore a good chance that they will not be in the same set in the future. This hypothesis is supported with good experimental results. We have also built our algorithm on the assumption that nodes with a smaller number of influential neighbours are less influential themselves. Hence, there is a relatively small probability that the exclusion of less effective neighbours at each iteration will lead to a profound difference in the final results, and thus they are extracted from the set of nodes.

## 4 Experimental Results

In this section, we demonstrate the practical performance of our approach on several datasets collected from real-world social networks. Our method is evaluated based on both the Degroot opinion formation model and our opinion formation model, as discussed above. We use three variations of the Degroot model, as explained below. The reason for using the Degroot model in addition to our own opinion formation model is to prevent the assumption of a 'chicken and egg' problem that might arise due to the similarities between our model and an opinion leader-finding algorithm. In other words, we show that our model is not the only one to give good performance using the algorithm's outputs[8]. However, our own opinion formation model shows good performance for our algorithm, and in a real-world situation in which the 'chicken and egg' problem is not a concern, our model and opinion-finding algorithm can be used simultaneously. The statistics of all datasets are summarised in Table 2, in which the parameters $N$ and $M$ denote the number of nodes and edges in the network, $k$ is the average degree of the nodes in the network, and $k_{max}$ is the maximum degree of a node that exists in the network. The parameter $C$ represents the average of the clustering coefficient of the nodes in the graph, and $LCC$ represents the size of the largest connected component in each graph. The Advogato network is an online community platform for developers of free software, and was launched in 1999. Nodes are users of Advogato and the directed edges represent trust relationships. In the Epinions network, nodes are the users and directed edges indicate trust relationships between users. Epinions is a trust network from an online social network. The Pokec network is a friendship network from the Slovak social network Pokec, in which nodes are users of Pokec and directed edges represent friendships.

Twitter is a directed network containing information about who follows whom on the platform. Nodes represent users and an edge shows that the left-hand user follows the right-hand one.
Firstly, to test the influence of the detected influential nodes in the opinion formation model, a random value in the range [0, 1] is assigned to each node to represent its opinion on certain subjects. If this value is close to zero, this indicates that the node has a negative view on that subject; conversely, if this value is close to one, this indicates that this node fully supports this subject. In all our experiments, we consider 10% of the nodes to be influential, and thus their opinions are set to one. We update each node's opinion based on the opinion formation model used in the experiment until the results converge to certain values and no further changes are seen. The final opinion in the network, which is the average of all nodes' opinions, is used as a benchmark to evaluate the influence of influential nodes.

---

[8] The datasets can be downloaded from: http://konect.uni-koblenz.de/networks/.

We conduct these tests for three different scenarios: (i) all initial values are set to 0; (ii) all initial values are set to 0.5; and (iii) the initial values are random, i.e. the values are set to a random number between zero and one. The value for parameter $\alpha$ is set to 0.8, since the algorithm works best for this value, as observed in Section 3.4. Due to space limitations we report only on the experiments with initial values of zero. The remainder of the experimental results are reported in the Appendix.

## 4.1 Test results for the proposed opinion formation model

In this section, the efficiency of the proposed method using the opinion formation model is compared to other commonly used methods, namely betweenness centrality, closeness centrality, eigenvector centrality and PageRank centrality. We assign initial values as explained at the beginning of Section 4. We run the test for our proposed opinion formation model and compare the values of the opinions for each method at each iteration. Thus, the results presented here are the averages of 20 runs of the simulation.

In these experiments, the opinions of the users were numbers in the range 0–1; if the value is close to one, then it means that a particular user supports the idea completely; its closeness to zero indicates the opposite. The y-axis in the figures denotes the average of the opinions of all nodes, and the x-axis indicates the time step in the test.

## 4.2 Test results for the Degroot opinion formation model

As in the previous section, we test our proposed method and compare the results to other methods; however, this time our tests are conducted for the Degroot model. All other evaluation criteria are the same as in Section 4.1.

## 4.3 Test results for three different variations of centrality for the Degroot opinion formation model

In this section, we compare different choices for $C_i$ (here we use $C_i$ instead of $T_{ij}$, which was used in the original Degroot model) in Equation 12 for the Degroot opinion formation model. The Degroot model is used for all of these tests. The three variations of $C_i$ are as follows: (i) using global centrality, as in Sections 4.1 and 4.2; (ii) including nodes that are two hops away in computing $C_i$ with a discount factor of 0.5; and (iii) using degree imbalances[9] without adding the neighbours' degree imbalances.

---

[9] Absolute values of the in-degree minus the out-degree of nodes.

Table 3

*Properties of test networks*

| Network | N | M | $<k>$ | $k_{max}$ | C | LCC |
|---|---|---|---|---|---|---|
| Advogato | 6,541 | 51,127 | 15.633 | 943 | 9.22 | 5,042 |
| Epinions | 75,879 | 508,837 | 13.412 | 3,079 | 6.57 | 75,877 |
| Twitter | 465,017 | 834,797 | 3.5904 | 678 | 0.0613 | 465,017 |
| Pokec | 1,632,803 | 30,622,564 | 37.509 | 20,518 | 4.68 | 1,632,803 |

**Algorithm 1** Mining Top *k* Most Influential Nodes

1: **input:** network $G = (V, E)$, size of result $k$
2: **output:** Top-$k$ influential nodes;
3: **For** each vertex $i$ **do**
4:     $influ_i^0 \leftarrow outdegree(i) - indegree(i)$
5: **end for**
6: $t \leftarrow 0$
7: **while** $size(V) > k$ **do**
8:     $t \leftarrow t + 1$
9:     **for** $i \leftarrow 1$ **to** $size(V)$ **do**
10:         $influ_i^{t+1} \leftarrow \alpha\, influ_i^0 + (1 - \alpha) \sum_{j \text{ is in-neighbors of } i} w_{ij}\, influ_j^t$
11:     **end for**
12:     $kth\ max^t \leftarrow kth\ largest\ element\ in\ influ_{v \in V}^t$
13:     $influ_{max}^t \leftarrow max_{v \in V}(influ_v^t)$
14:     $influ_{min}^t \leftarrow min_{v \in V}(influ_v^t)$
15:     **for each vertex** $v$ **in** $V$ **do**
16:         **if** $kth\ max^t - influ_v^{(t)} > 2(1 - \alpha)^t (influ_{max}^t - influ_{min}^t)$ **then**
17:             $V \leftarrow V - v$
18:         **end if**
19:     **end for**
20: **end while**
21: **return** $V$

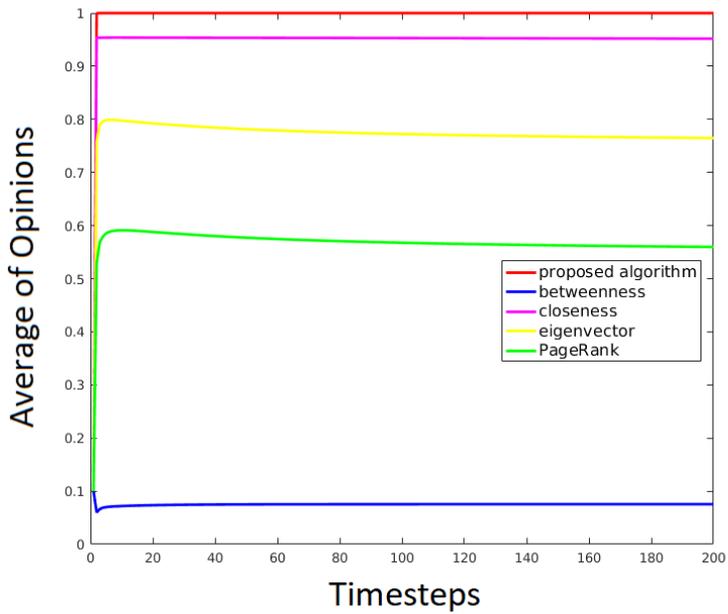

(a) Twitter Network

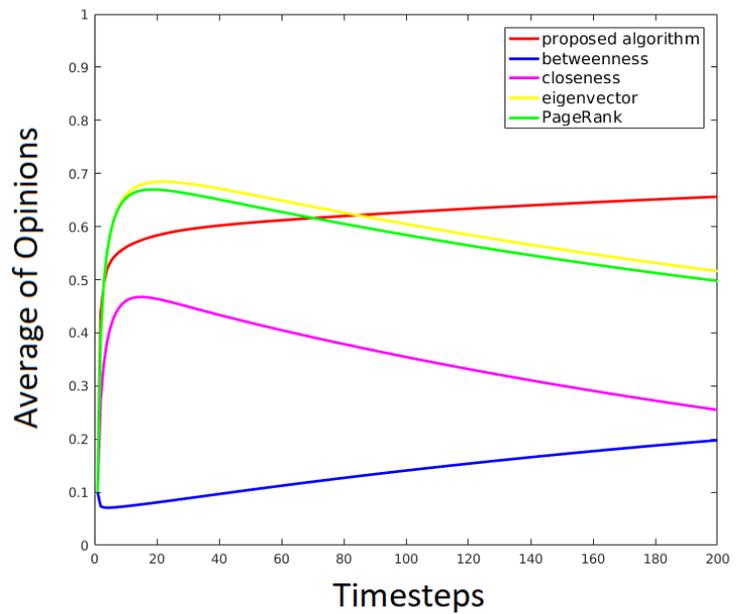

(b) Pokec Network

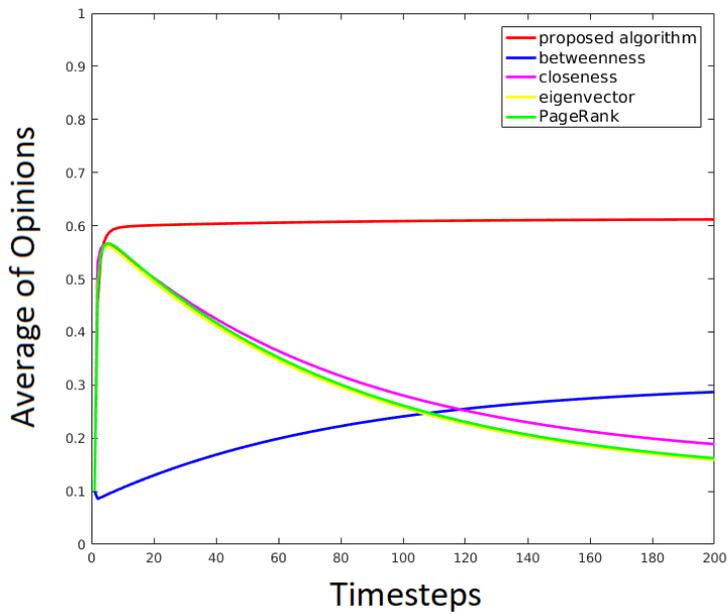

(c) Advogato Network

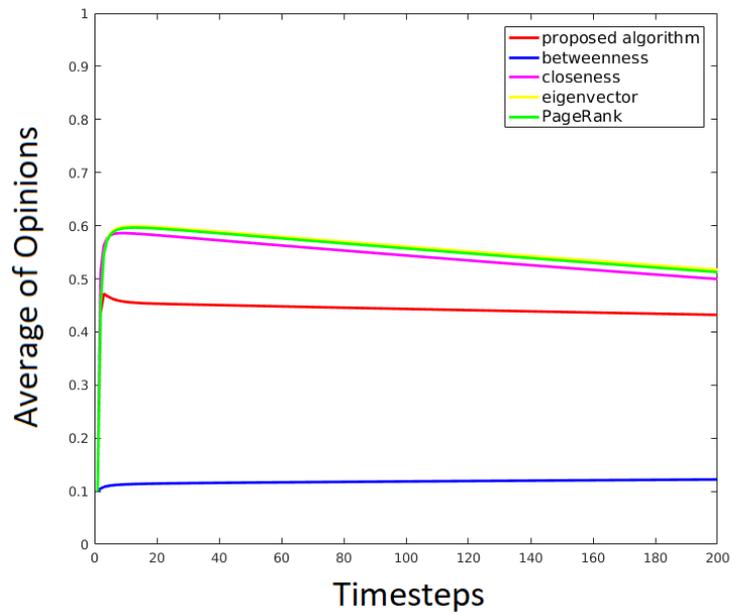

(d) Epinions Network

*Figure 2. Comparison of methods for finding opinion leaders in the Degroot opinion formation model with global centrality and initial values of zero*

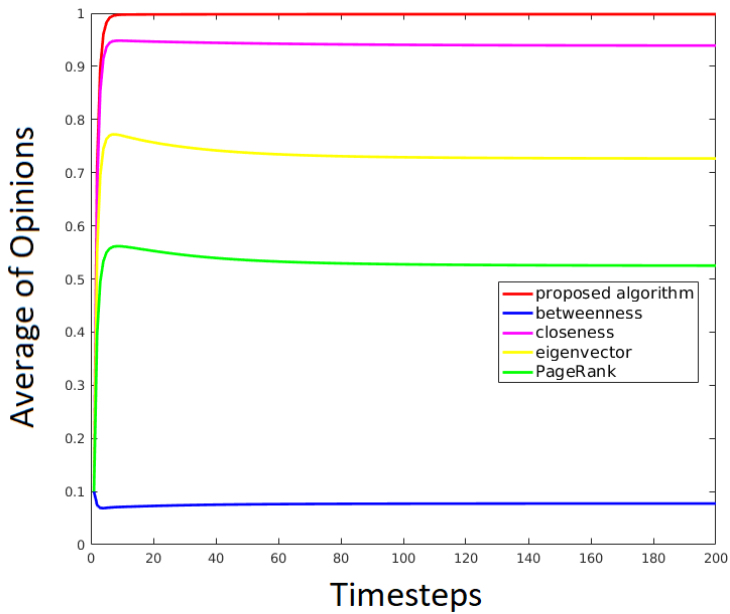

(a) Twitter Network

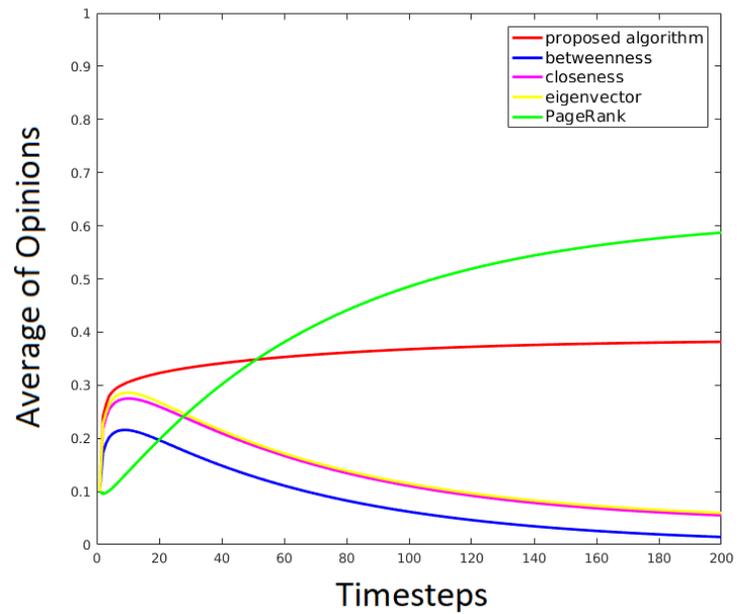

(b) Pokec Network

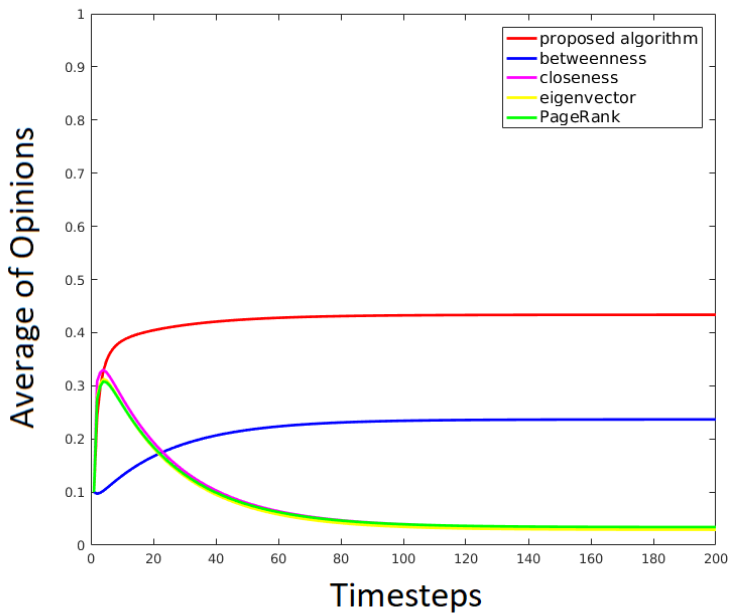

(c) Advogato Network

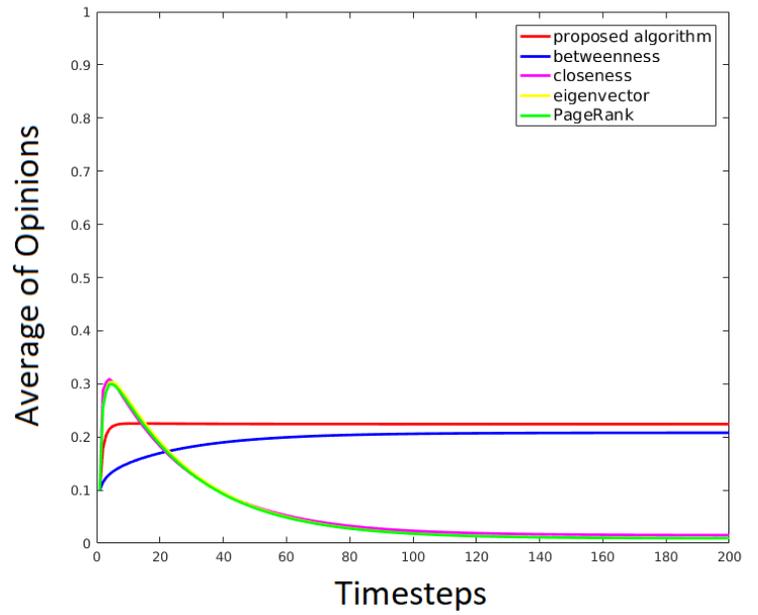

(d) Epinions Network

*Figure 3. Comparison of methods for finding opinion leaders in the Degroot opinion formation model with two-hop centrality and initial values of zero*

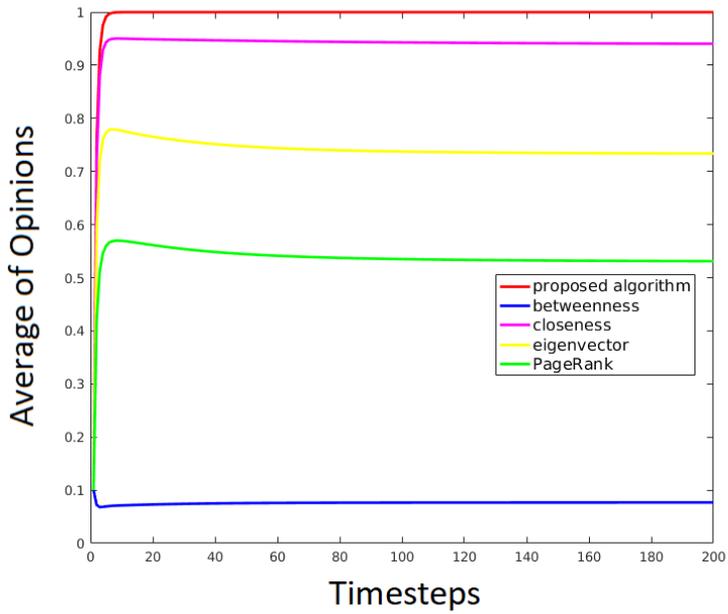
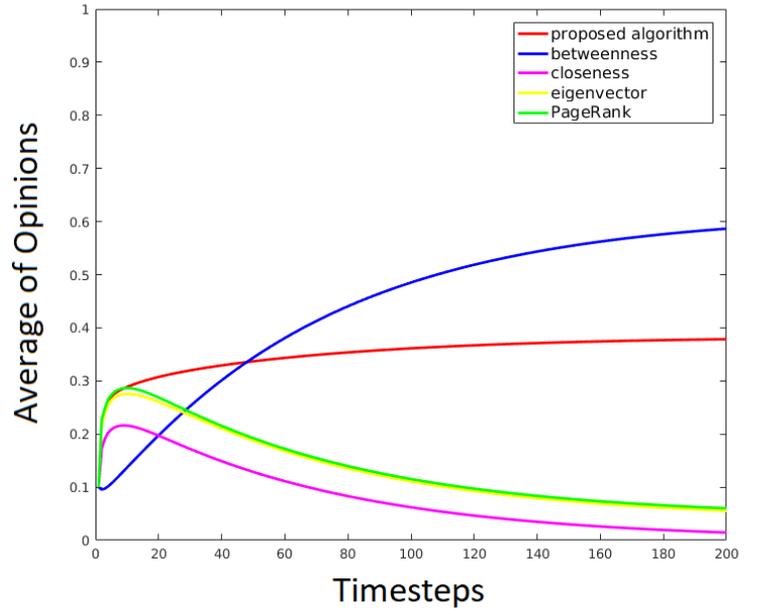
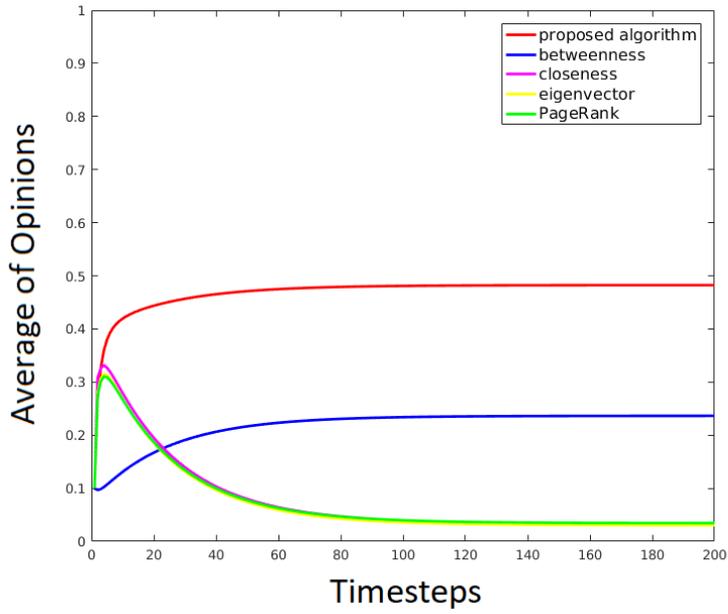
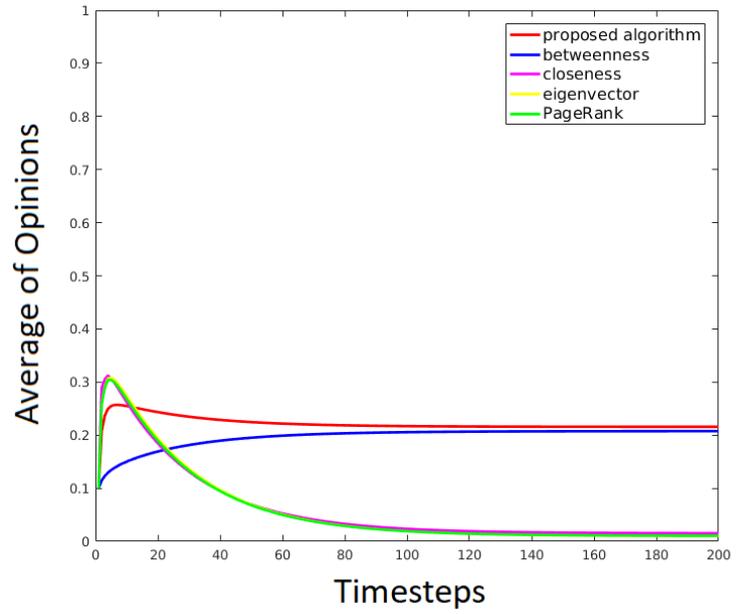

*Figure 4. Comparison of methods for finding opinion leaders in the Degroot opinion formation model with degree imbalance as centrality with initial values of zero*

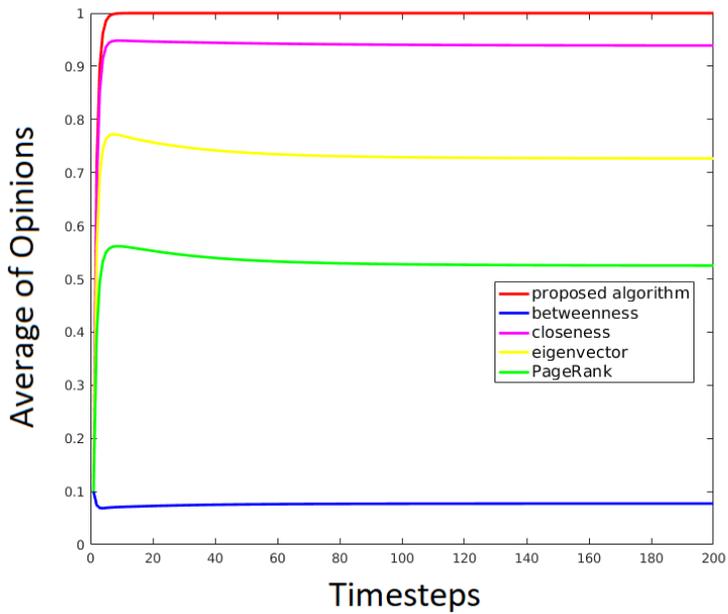
(a) Twitter Network

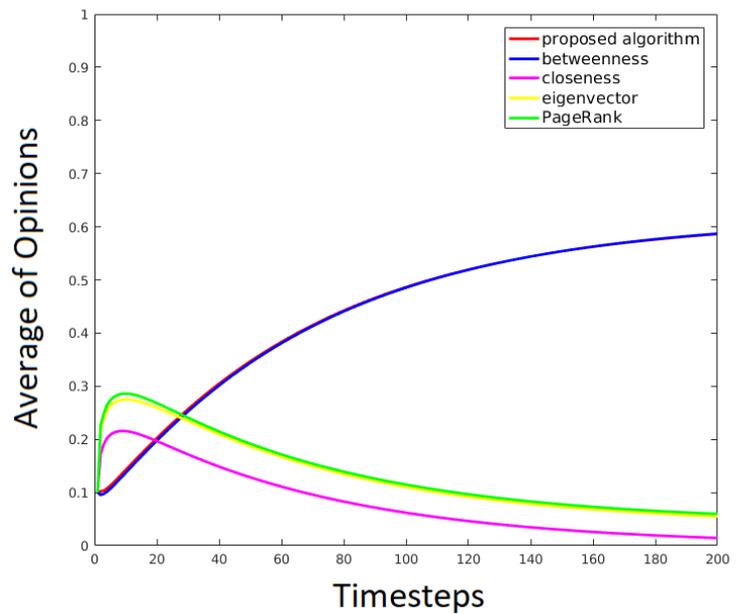
(b) Pokec Network

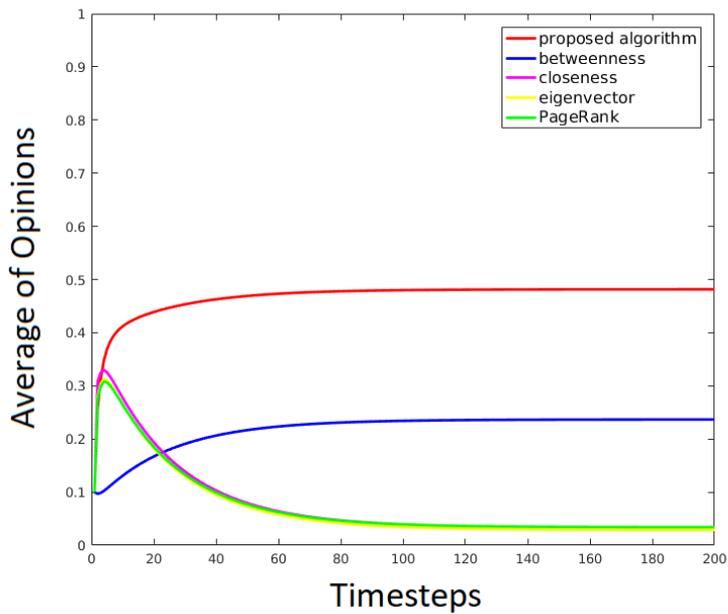
(c) Advogato Network

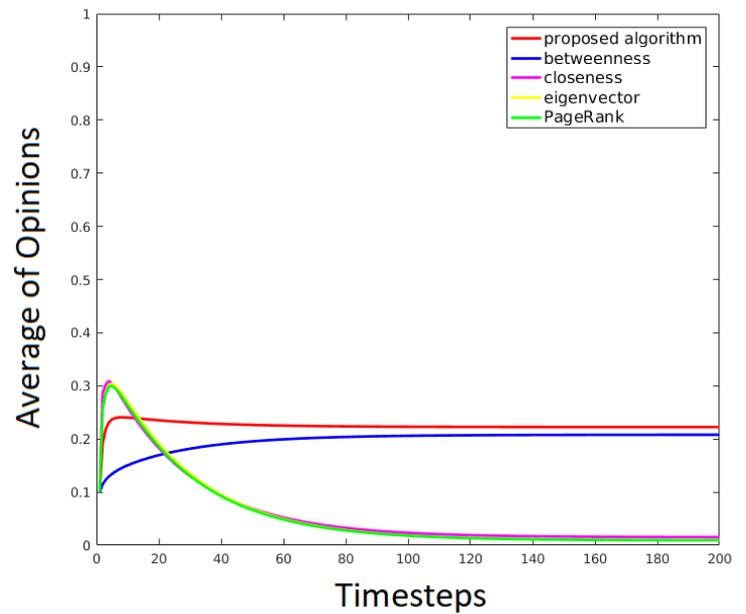
(d) Epinions Network

*Figure 5. Comparison of methods for finding opinion leaders in the proposed opinion formation model with initial values of zero*

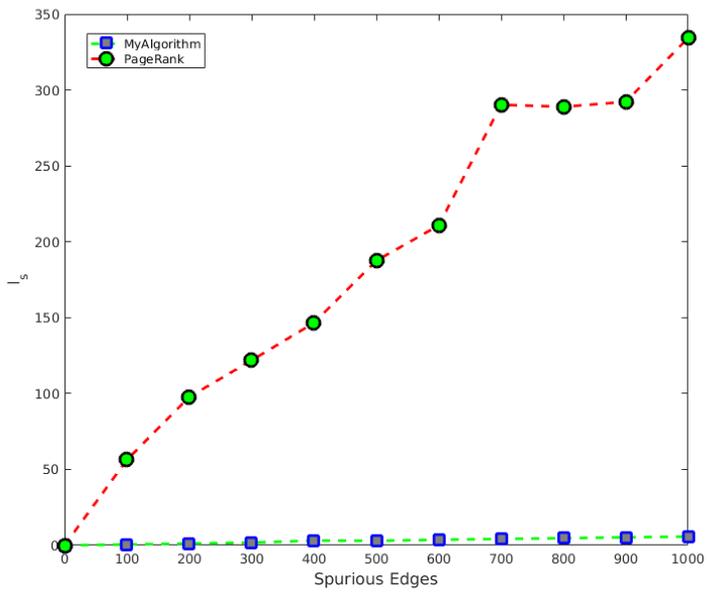

(a) Twitter Network

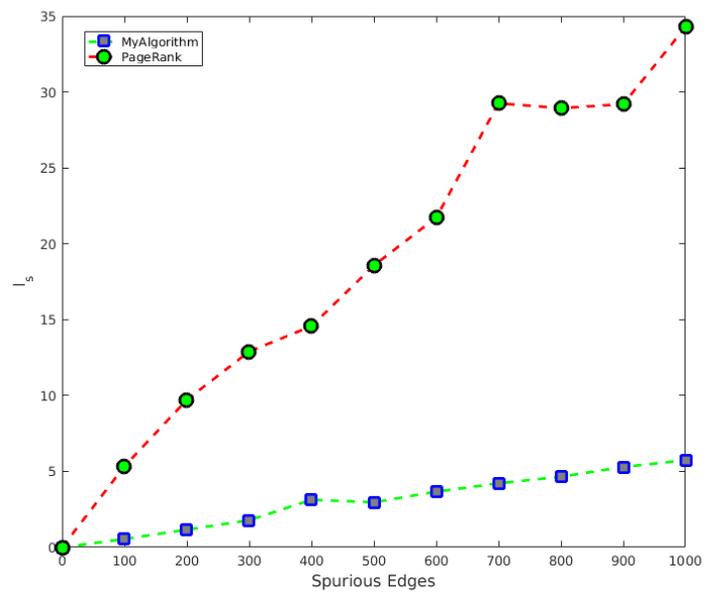

(b) Pokec Network

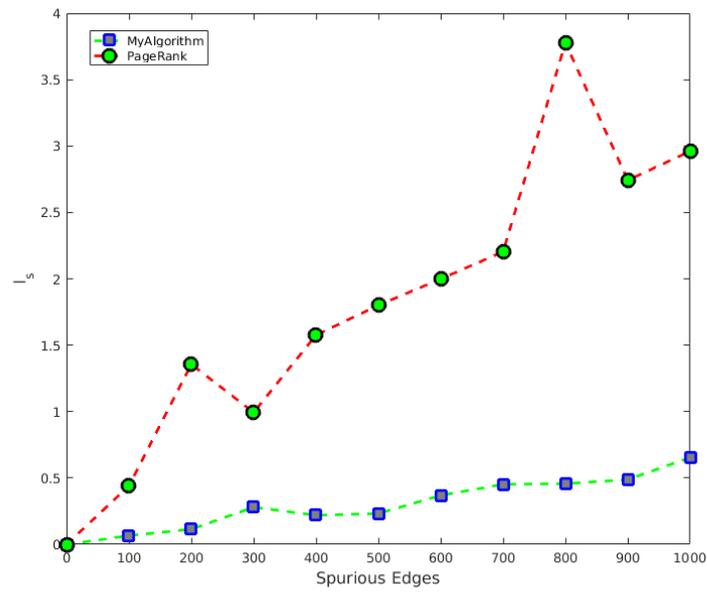

(c) Advogato Network

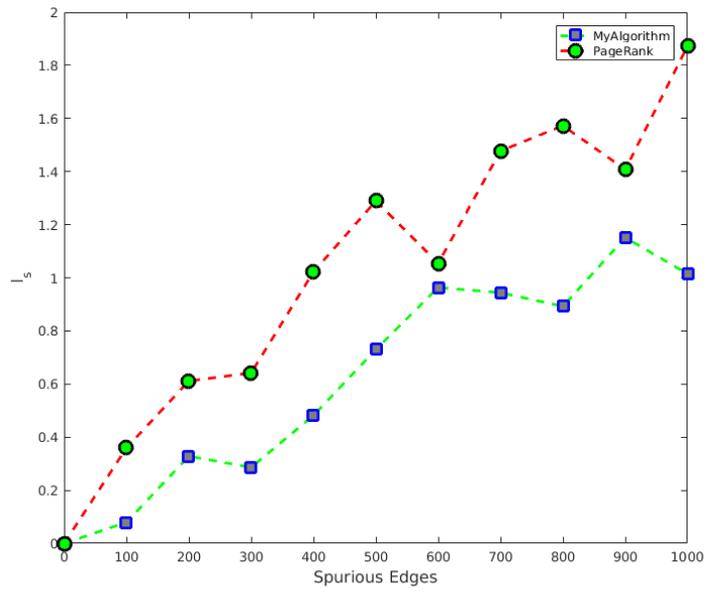

(d) Epinions Network

*Figure 6. Comparison of fault tolerance in the proposed method and the PageRank method*

## 4.4 Evaluation of results

As we can observe from Figure 2 and from all the datasets except the Epinions network, the values of the opinions increase and finally converge to a particular value. In the Epinions dataset, the values of the opinions decrease but also converge as in the other datasets. We can see the same behaviour in Figures 3 and 4 in all datasets (except for the Epinions network in Figure 3). In all cases, the proposed method was superior to the other methods of finding opinion leaders, except for the case of the Epinions network in Figure 2 and the Pokec network in Figure 4. As can be seen from Figure 5, our method outperforms the other methods of finding opinion leaders since it considers both influence and conformity simultaneously. In the Pokec network, the proposed method gives the same performance as the betweenness centrality and is better than the other choices.

Another interesting observation is that there is no significant difference between the curves except for Pokec network in Figures 2, 3 and 4 and the Epinions network in Figure 2. Both the results for the proposed opinion formation model and for the Degroot model converge immediately to almost the same curves, regardless of the method used. This is due to the fact that if the initial opinion leaders are chosen correctly, then there is no significant difference of what opinion formation model they have and this will converge to the same results.

## 4.5 Fault tolerance against noisy data

Fault tolerance is of paramount importance, especially in situations where some of the links in the input data do not exist in the real world and our data contain errors. Social networking data may be associated with uncertainty, especially in cases where we determine relationships between agents. For example, in a protein network obtained from a biological experiment, there is always the possibility of error when modelling the links of the network. In order to achieve the task of finding the top $k$ influential nodes in these networks, we require algorithms with great tolerance against such errors. In this section, we examine the fault tolerance of our proposed method against the PageRank algorithm. In order to create a graph containing errors, we add some edges to our original graph. The addition of wrong connections to the graph makes the graph 'noisy'. In the next step, we compute the centrality for this noisy graph and measure the error between the noisy and normal graphs using the following equation:

$$I_s = \sum_{i=1}^{N} \left| S_i' - S_i \right| \qquad (49)$$

$S_i'$ and $S_i$ correspond to the centrality score of node $i$ in the noisy and original graphs, respectively. In Figure 9, the term 'spurious edges' refers to the number of edges that do not actually exist in the graph. As previously stated, these are errors in the network. As shown in the same figure, our centrality is more robust to noisy data than the PageRank centrality, as indicated by its lower value of $I_s$.

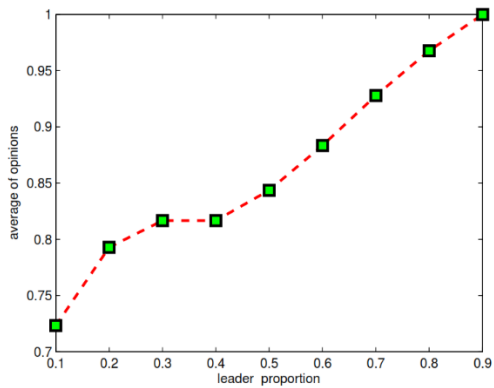

*Figure 7. Effect of the proportion of leader nodes on the average of the final opinions of nodes in the Pokec network*

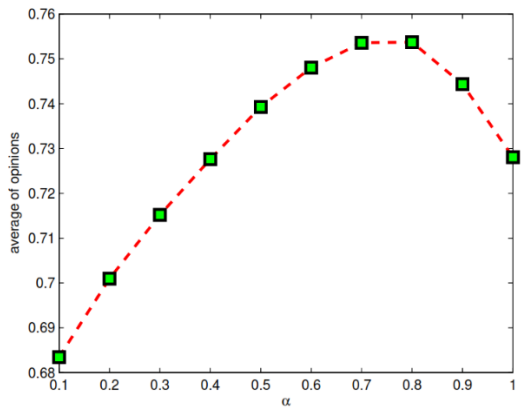

*Figure 8. Effect of different values of α on the average of the final opinions of nodes in the Pokec network*

Figure 7 shows the final opinions in the network that are reached by multiple proportions of leaders in the Pokec network. This curve has an ascending shape, and by increasing the number of leaders, the average of the final opinion of the system increases. The results for the other networks tested are similar. In Figure 8, we can observe the effects of different values of $\alpha$ on the average of the opinions of the nodes in the Pokec network. A larger value of $\alpha$ makes leaders more local, while a smaller value makes them more global. The results for the other networks tested are similar.

## 5 Conclusion and future works

In this research paper, we propose a new centrality measure to find opinion leaders and an algorithm to find these leaders. Furthermore, we introduce a new opinion formation model to calculate the opinions of the network nodes.

The main advantage of our method is that it considers both the input and output links, while previous methods have mostly focused on the output links and have ignored the role of the input links. We test our proposed opinion-leader-finding algorithm with our proposed opinion formation model and the Degroot opinion formation model. Our method outperforms classic centrality measures, and also has the advantage of fault tolerance.

Directions for future research may include the implementation of our method in a distributed framework (e.g. Hadoop, using the MapReduce programming model). Another interesting area of research would be the application of these methods to graphs for which we do not have full knowledge about the network's structure, or to graphs in which the topology of the graph changes constantly, thus giving the network a dynamic structure.